\documentclass[10pt,conference]{IEEEtran}
\IEEEoverridecommandlockouts
\usepackage{cite}
\usepackage{amsmath,amssymb,amsfonts}
\usepackage[noend]{algorithmic}
\usepackage{graphicx}
\usepackage{textcomp}
\usepackage{xcolor}
\usepackage{amssymb}
\usepackage{indentfirst}
\usepackage{amsmath}
\usepackage{autobreak}
\usepackage{bm}
\usepackage{booktabs}
\usepackage[font=small]{caption}
\usepackage{subcaption}

\usepackage{tikz}
\usepackage{pgfplots}
\usetikzlibrary{plotmarks}
\usetikzlibrary{matrix,positioning,shapes,arrows,decorations,calc,fit}
\usetikzlibrary{decorations.pathreplacing}
\usetikzlibrary{spy,backgrounds}
\usetikzlibrary{external}
\usetikzlibrary{patterns}
\usetikzlibrary{shapes,arrows}
\usetikzlibrary{shadows.blur}
\usetikzlibrary{shapes.symbols}
\usetikzlibrary{
	pgfplots.groupplots,
	matrix
}

\usepackage{epsfig}
\usepackage{setspace}
\usepackage{algorithm}
\usepackage{changepage}
\usepackage{amsthm} 
\usepackage{soul,color}
\usepackage{hyperref}
\usepackage{lipsum}
\usepackage{cuted}
\usepackage{multirow}
\theoremstyle{definition}

\IEEEaftertitletext{\vspace{-1\baselineskip}}

\begin{document}

\setlength{\abovedisplayskip}{3pt}
\setlength{\belowdisplayskip}{3pt}

\title{Ordered-Statistics Decoding with Adaptive Gaussian Elimination Reduction for Short Codes}
\author{\IEEEauthorblockN{
       Chentao Yue, Mahyar Shirvanimoghaddam, Branka Vucetic, and Yonghui Li}
\IEEEauthorblockA{School of Electrical and Information Engineering, The University of Sydney, NSW, Australia \\
\{chentao.yue, mahyar.shm, branka.vucetic, yonghui.li\}@sydney.edu.au}

}

\maketitle

\begin{abstract}
In this paper, we propose an efficient ordered-statistics decoding (OSD) algorithm with an adaptive Gaussian elimination (GE) reduction technique. The proposed decoder utilizes two decoding conditions to adaptively remove GE in OSD. The first condition determines whether GE could be skipped in the OSD process by estimating the decoding error probability. Then, the second condition is utilized to identify the correct decoding result during the decoding process without GE. The proposed decoder can break the ``complexity floor'' in OSD decoders introduced by the GE overhead. Simulation results advise that when compared with the latest schemes in the literature, the proposed approach can significantly reduce the decoding complexity at high SNRs without any degradation in the error-correction capability.

\end{abstract}

\IEEEpeerreviewmaketitle

\vspace{-0.3em}
\section{Introduction}
\vspace{-0.3em}

   Ultra-reliable and low-latency communications (URLLC) have attracted a lot of attention in 5G and the upcoming 6G for mission-critical services\cite{3GPPRelease16,Mahyar2019ShortCode,tataria20216g}. URLLC is renowned by its requirements for significantly lower latency compared to 4G long term evolution (4G-LTE) and high transmission reliability requiring the block error rate (BLER) lower than $10^{-4}$. These stringent requirements necessitate the use of low-complexity decoders and short block-length codes ($\leq 150$ bits) in coding scheme design \cite{3GPPRelease16,Mahyar2019ShortCode}. Additionally, URLLC requires bit-level granularity in the block lengths and coding rates, to accommodate scenarios with varying latency and bandwidth constraints \cite{Mahyar2019ShortCode}, which complicates the coding system further.
   
    Main candidates of short block-length codes for URLLC have been thoroughly reviewed in \cite{Mahyar2019ShortCode,liva2016codeSurvey}. As has been shown, short Bose-Chaudhuri-Hocquenghem (BCH) codes and cyclic-redundancy-check-aided Polar (CRC-Polar) codes have superior BLER performance that is close to the normal approximation (NA) bound\cite{Mahyar2019ShortCode,erseghe2016coding}. However, it is challenging for BCH and CRC-Polar to provide bit-level granularity with the optimal error-correction capability, as they are originally available at certain block-lengths and rates \cite{lin2004ECC,Designofpolar}, while the best known linear codes (BKLC) of different lengths and rates have different structures (see code table \cite{Grassl:codetables}). Recently, \cite{papadopoulou2021short} used the genetic algorithm to design high-performance short codes by evolving random codes. \cite{yue2022efficient} shows that near-optimal rate-compatible codes can be constructed by selecting random generator matrix columns and efficiently decoded by universal decoders. Therefore, universal decoders are favored in URLLC \cite{yue2022efficient}, as they can decode any linear block codes, including BCH, CRC-Polar codes, and BKLC, as well as their rate-compatible versions. 
    
    Ordered-statistics decoding (OSD) \cite{Fossorier1995OSD} is a near-maximum-likelihood (near-ML) universal decoder that rekindles interests recently \cite{yue2021probability,Chentao2019SDD,NewOSD-5GNR,yue2021linear,wang2021self,wang2021efficient,choi2019fast}. It can decode any linear block code with near-ML BLER performance. OSD has two main phases, namely, \textit{preprocessing} and \textit{reprocessing}. In preprocessing, it sorts the received codeword bits according to their reliabilities (by which high reliable bits are distinguished), and permutes the columns of the code generator matrix accordingly. The permuted matrix is then transformed into the systematic form by Gaussian elimination (GE), where the information set is associated with high reliable bits. In reprocessing, a number of test error patterns (TEPs) are attempted to flip the high reliable bits. The remaining low reliable bits are recovered by \textit{re-encoding} the flipped high reliable bits with the systematic permuted generator matrix.
    
    Let $C$ be the notation of complexity. The average decoding complexity of OSD is roughly characterized as
    \begin{equation} \label{equ::CmpOSD}
        C_{\mathrm{OSD}} =  C_{\mathrm{Preprocessing}} + N_a  C_{\mathrm{Re-encoding}},
    \end{equation}
    where $N_a$ is the average number of TEPs processed in decoding a single codeword. Recent years have seen many works towards reducing $C_{\mathrm{OSD}}$ by reducing the number $N_a$ \cite{yue2021probability,Chentao2019SDD,NewOSD-5GNR,yue2021linear,Wu2007OSDMRB,FossorierBoxandMatch,choi2019fast, jin2006probabilisticConditions,WJin2007MultipleBiases,wang2021efficient}. For example, \cite{yue2021probability,Chentao2019SDD,Wu2007OSDMRB,choi2019fast} proposed techniques to identify unpromising TEPs that can be discarded without processing, and \cite{yue2021probability,Wu2007OSDMRB, jin2006probabilisticConditions} designed approaches to terminate OSD early rather than processing all possible TEPs. These approaches can reduce $N_a$ to a very low level at high signal-to-noise ratios (SNRs). In this situation, $C_{\mathrm{OSD}}$ will be dominated by preprocessing rather than reprocessing. Specifically, GE has the complexity as high as $O(nk^2)$ for a $k\times n$ generator matrix \cite[Table I]{Fossorier1995OSD}. As a result, when $N_a$ is small, GE introduces a \textit{complexity floor} to OSD. That is, a complexity component hardly reducible \cite{yue2022efficient}. This complexity floor hinders the application of OSD in URLLC, especially for high-SNR scenarios. Recently, \cite{choi2021fast} proposed using multiple offline produced generator matrices to replace GE in OSD. Nevertheless, this could introduce extra overheads at low-to-moderate SNRs, as it performs several reprocessings over multiple generator matrices. Until now, the complexity floor of OSD has not been fully addressed in the literature. 
    
    In this paper, we design an OSD decoder with adaptive GE reduction. Specifically, the decoder will skip GE if doing so has a negligible impact on the decoding error performance. To accomplish this, two decoding conditions are developed. The first condition decides whether to skip GE by evaluating BLER performance with and without GE. If GE is skipped, the re-encoding process will be performed with the original generator matrix of the code. The second condition determines whether the correct decoding result has been found in the decoding process without GE. If so, decoding is terminated early to reduce complexity; if not, the standard OSD is performed as a supplement to prevent BLER from degrading. Our verification shows that at high SNRs, the proposed decoder can avoid GE in almost all decoding attempts with nearly the same BLER performance as the standard OSD. Owing to the effective GE reduction, the proposed decoder achieves a significantly lowered complexity at high SNRs compared to the latest OSD approaches from the literature \cite{choi2021fast,yue2021probability}.
    
    The rest of this paper is organized as follows. Section \ref{sec::Preliminaries} presents the preliminaries. Section \ref{sec::Algorithm} describes the proposed decoding algorithm. Section \ref{sec::Performance and Complexity} analyzes the decoding error performance and complexity. Section \ref{Sec::Simulation} verifies the BLER and complexity of the proposed decoder via simulations. Finally, Section \ref{sec::Conclusion} concludes the paper.
    
    \emph{Notation}: We use $[a]_u^v = [a_u,\ldots,a_v]$ to denote a row vector containing element $a_{\ell}$ for $u\le \ell\le v$. For simplicity, we do not distinguish random variables and their samples throughout the paper, while possible abuse of notations will be specified.

\vspace{-0.1em} 
\section{Preliminaries} \label{sec::Preliminaries}
\vspace{-0.1em} 
Let $\mathcal{C}(n,k)$ denote a binary linear block code, where $n$ and $k$ are the lengths of the codeword and the information block, respectively. $\mathcal{C}(n,k)$ is defined by its generator matrix $\mathbf{G}$; that is, an information sequence $\mathbf{b} =  [b]_1^k$ is uniquely encoded to a codeword $\mathbf{c} = [c]_1^n$ by $\mathbf{c} = \mathbf{b}\mathbf{G}$. In this paper, we assume that $\mathbf{G}$ is systematic, i.e.,  $\mathbf{G} =  [\mathbf{I}_k \  \mathbf{P}]$, where $\mathbf{I}_k$ is a $k\times k$ identity matrix and $\mathbf{P}$ is the parity sub-matrix.

We consider an additive white Gaussian Noise (AWGN) channel and binary phase shift keying (BPSK) modulation. Let $\mathbf{s} = [s]_1^n$ denote the modulated signals, where $s_{i} = (-1)^{c_{i}}\in \{\pm 1\}$. At the channel output, the received signal is given by $\mathbf{r} = \mathbf{s} + \mathbf{w}$, where $\mathbf{w}$ is the AWGN vector with zero mean and variance $N_{0}/2$, for $N_0$ being the single-band noise power spectrum density. SNR is accordingly defined as $\mathrm{SNR} = 2/N_0$. Without loss of generality, approaches provided in this paper can be extended to other modulation schemes.

At the receiver, the bitwise hard-decision estimate $\mathbf{y}= [y]_{1}^n$ of codeword $\mathbf{c}$ is obtained according to: $y_{i} = 1 $ for $r_{i}<0$ and $y_{i} = 0$ for $r_{i}\geq 0$. If codewords in $\mathcal{C}(n,k)$ have equal transmission probabilities, the log-likelihood ratio (LLR) of the $i$-th received symbol is defined as ${\ell}_{i} \triangleq \ln \frac{\mathrm{Pr}(c_{i}=0|r_{i})}{\mathrm{Pr}(c_{i}=1|r_{i})}$, which is further simplified to ${\ell}_{i} = \frac{4r_{i}}{N_{0}}$ with BPSK \cite{lin2004ECC}. Thus, we define $\alpha_{i} \triangleq |r_{i}|$ (the scaled magnitude of LLR) as the reliability of $y_i$, where $|\cdot|$ is the absolute operation.

OSD preprocesses received signals according to reliabilities. First, a permutation $\pi_{1}$ is performed  to sort the reliabilities $\bm{\alpha} = [\alpha]_1^n$ in descending order, and the ordered reliabilities are obtained as $\pi_1(\bm{\alpha})$. Then, the generator matrix is accordingly permuted (in terms of columns) to $\pi_1(\mathbf{G})$. Next, OSD obtains the systematic form of $\pi_1(\mathbf{G})$ as $\widetilde{\mathbf{G}} =  [\mathbf{I}_k \  \widetilde{\mathbf{P}}]$ by performing GE. We represent the GE operation as $\widetilde{\mathbf{G}} = \mathbf{E}(\pi_2(\pi_1(\mathbf{G})))$, where $\mathbf{E}$ (dimension $k\times k$) represents row operations, and $\pi_{2}$ represents an additional column permutation which ensures that the first $k$ columns of $\pi_1(\mathbf{G})$ are linearly independent. Accordingly, $\mathbf{r}$, $\mathbf{y}$, and $\bm{\alpha}$, are permuted into $\widetilde{\mathbf{r}} = \pi_2(\pi_1(\mathbf{r}))$, $\widetilde{\mathbf{y}} = \pi_2(\pi_1(\mathbf{y}))$, $\widetilde{\bm{\alpha}} = \pi_2(\pi_1(\bm{\alpha}))$, respectively.

Since $\pi_2$ only marginally disrupts the descending order of $\widetilde{\bm{\alpha}}$ \cite[Eq. (59)]{Fossorier1995OSD}, the first $k$ positions of $\widetilde{\mathbf{y}}$, denoted by $\widetilde{\mathbf{y}}_{\mathrm{B}} =[\widetilde{y}]_1^k$, are referred to as the most reliable basis (MRB) \cite{Fossorier1995OSD}. To eliminate errors in MRB, a length-$k$ TEP $\mathbf{e} = [e]_1^k$ is added to $\widetilde{\mathbf{y}}_{\mathrm{B}}$ to obtain a codeword estimate by re-encoding, i.e., $\widetilde{\mathbf{c}}_{\mathbf{e}} = \left(\widetilde{\mathbf{y}}_{\mathrm{B}}\oplus \mathbf{e}\right)\widetilde{\mathbf{G}}$, where $\widetilde{\mathbf{c}}_{\mathbf{e}}$ is the ordered codeword estimate with respect to the TEP $\mathbf{e}$. In reprocessing, a list of TEPs are re-encoded to generate multiple codeword candidates. The maximum Hamming weight of TEP is limited by a parameter $m$, namely, decoding order of OSD. Thus, the number of TEPs processed in an order-$m$ OSD is up to $\sum_{i=0}^{m}\binom{k}{i}$. For a code with the minimum Hamming weight $d_{\mathrm{H}}$, OSD with order $m = \lceil d_{\mathrm{H}}/4-1\rceil$ achieves the ML decoding performance \cite{Fossorier1995OSD}.

With BPSK, the best ordered codeword estimate $\widetilde{\mathbf{c}}_{\mathrm{best}}$ is found by minimizing the weighted Hamming distance between codeword estimate $\widetilde{\mathbf{c}}_{\mathbf{e}}$ and $\widetilde{\mathbf{y}}$, which is defined as \cite{Wu2007OSDMRB}
    	\begin{equation} \small \label{equ::Prelim::WHD_define}
    		 \mathcal{D}(\widetilde{\mathbf{c}}_{\mathbf{e}},\widetilde{\mathbf{y}}) \triangleq \sum_{1 \leq i \leq n } (\widetilde{c}_{\mathbf{e},i} \oplus \widetilde{y}_{i}) \widetilde{\alpha}_{i}.
    	\end{equation}
Finally, the decoding result $\hat{\mathbf{c}}_{\mathrm{best}}$ is output by performing inverse permutations over $\widetilde{\mathbf{c}}_{\mathrm{best}}$, i.e., $\hat{\mathbf{c}}_{\mathrm{best}} = \pi_1^{-1}(\pi_2^{-1}(\widetilde{\mathbf{c}}_{\mathrm{best}}))$.

\vspace{-0.25em} 
\section{OSD with Adaptive GE Reduction}   \label{sec::Algorithm}
\vspace{-0.25em} 
\subsection{Overall Description}
\vspace{-0.25em} 
     \begin{figure} 
     \centering
        \definecolor{mycolor1}{rgb}{0.00000,0.44706,0.74118}%
        \definecolor{mycolor2}{rgb}{0.00000,0.44700,0.74100}%
        \tikzstyle{terminator} = [rectangle, draw, text centered, rounded corners, minimum height=2em]
        \tikzstyle{process} = [rectangle, draw, text centered, minimum height=2em]
        \tikzstyle{decision} = [diamond, draw, text centered, minimum height=1em,aspect=2.5]
        \tikzstyle{connector} = [draw, -latex']
        \begin{tikzpicture}[node distance=2cm]
        \node at (-2,0) [terminator] (start) {\footnotesize Start Decoding};
        \node [decision] at (-2,-1.3) (con1) {\footnotesize Condition 1};
        \node [process] at (2,-1.3) (NonGE) {\footnotesize Non-GE OSD (order $m\!-\!1$)};
        \node [decision] at (2,-2.6) (con2) {\footnotesize Condition 2};
        \node [process] at (-2,-2.6) (GE) {\footnotesize Standard OSD (order $m$)};
        \node at (0,-3.9) [terminator] (end) {\footnotesize Finish Decoding};
        \path [connector] (start) -- (con1);
        \path [connector] (con1) -- (NonGE);
        \path [connector] (con1) -- (GE);
        \path [connector] (NonGE) -- (con2);
        \path [connector] (con2) -- (GE);
        \path [connector] (con2) |- (end);
        \path [connector] (GE) |- (end);
        
        \node[draw=none] at (-1.6, -2.0) (No) {\footnotesize No};
        \node[draw=none] at (-0.4, -1.1) (Yes) {\footnotesize Yes};
        \node[draw=none] at (1.6, -3.3) (yes) {\footnotesize Yes};
        \node[draw=none] at (0.4, -2.4) (No) {\footnotesize No};
        
        \end{tikzpicture}
    	\vspace{-0em}
        \caption{The structure of the proposed decoder.}
    	\vspace{-0em}
    	\label{Fig::structure}
        
	\end{figure}
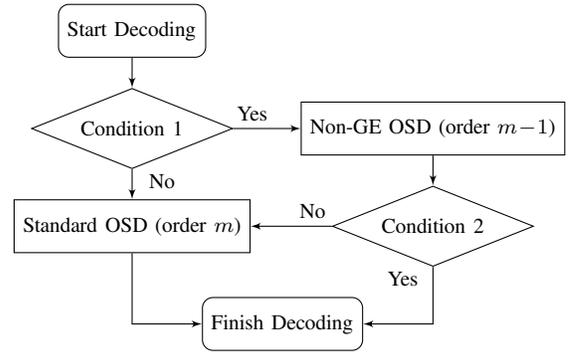

We proposed an OSD algorithm that can adaptively skip its reprocessing stage. When reprocessing is skipped, the original hard-decision estimate $\mathbf{y}$ and generator matrix $\mathbf{G}$ will be used for re-encoding (instead of using $\widetilde{\mathbf{y}}$ and $\widetilde{{\mathbf{G}}}$). Specifically, let $\mathbf{y}_{\mathrm{B}}$ denote the first $k$ positions of $\mathbf{y}$, i.e., $\mathbf{y}_{\mathrm{B}} = [y]_1^k$. Then, given a TEP $\mathbf{e}$, a codeword estimate is directly recovered by  $\mathbf{c}_{\mathbf{e}} = \left(\mathbf{y}_{\mathrm{B}}\oplus \mathbf{e}\right)\mathbf{G}$. Similar to the standard OSD, a list of TEPs will be used in re-encoding, and the maximum allowed Hamming weight of TEPs is limited by $m'$. Finally, if $\mathbf{c}_{\mathbf{e}}$ is identified as the best codeword estimate, it will be directly output as the decoding result with no inverse permutation required. We referred to such a decoding process without GE as the Non-GE OSD, and $m'$ is its decoding order. 

The structure of the proposed decoder is illustrated in Fig \ref{Fig::structure}. At the start of decoding, the decoder decides whether to conduct the Non-GE OSD or the standard OSD according to ``Condition 1''. If the Non-GE OSD is performed, ``Condition 2'' will determine whether the Non-GE OSD has found the correct decoding result. If not, the standard OSD will be conducted following the Non-GE OSD to avoid degraded decoding performance. We set that the Non-GE OSD is one-order lower than the standard OSD, i.e., $m'=\max(m-1,0)$. The reasons are that 1) if $m'\geq m$, the Non-GE OSD has a higher complexity than the standard OSD, which negates the need for GE reduction and worsens the worst-case decoding complexity, and 2) if $m'$ is too small, the Non-GE OSD may easily fail to find the correct result.

As seen, the design of ``Condition 1'' and ``Condition 2'' is of importance for the proposed decoder. We note that the standard OSD  in Fig. \ref{Fig::structure} can be implemented by any improved variants of OSD; for example, the efficient probability-based OSD (PB-OSD) proposed recently \cite{yue2021probability}.

\vspace{-0.25em} 
\subsection{The First Condition}
\vspace{-0.25em} 
To derive the first condition, let us first consider the BLER performance of OSD, which is represented as \cite[Eq. (69)]{Fossorier1995OSD}
\begin{equation}
    \mathrm{P_e} \leq (1 - \mathrm{P_{list}}) + \mathrm{P_{ML}},
\end{equation}
where $\mathrm{P_{ML}}$ is the ML BLER of $\mathcal{C}(n,k)$. $\mathrm{P_{ML}}$ is mainly characterized by the code structure and in particular, the code minimum Hamming weight $d_{\mathrm{H}}$. $\mathrm{P_{list}}$ is the probability that some TEP can eliminate the the errors over MRB, whose value depends on the decoding order $m$ \cite[Eq. (24)]{dhakal2016error}.

Let $\mathrm{P'_{list}}$ denote the probability that the Non-GE OSD can eliminate the errors over $\mathbf{y}_{\mathrm{B}}$ by some TEP. Thus, BLER of Non-GE OSD is upper bounded by $\mathrm{P_e} \leq (1 - \mathrm{P'_{list}}) + \mathrm{P_{ML}}$. Therefore, if $\mathrm{P'_{list}} =
 \mathrm{P_{list}}$, the Non-GE OSD will deliver the same BLER as OSD. In other words, the Non-GE OSD is sufficient to find the correct decoding result and thus GE is not necessary. $\mathrm{P'_{list}} =
 \mathrm{P_{list}}$ can be satisfied when SNR is asymptotically large (i.e., $N_0\to 0$). To see this, consider $\mathrm{P_{list}}$ derived from \cite[Lemma 1]{yue2021revisit}), i.e.,
 \begin{equation} \label{equ::Ana::Plist}
     \mathrm{P_{list}} = \sum_{i=0}^{m}\int_{x = 0}^{\infty} \binom{k}{i} (p(x))^{i}(1-p(x))^{k-i} f_{\widetilde{\alpha}_{k+1}}(x) dx,
 \end{equation}
 where $f_{\widetilde{\alpha}_{k+1}}(x)$ is the probability density function ($\mathrm{pdf}$) of the $(k+1)$-th ordered reliability, $\widetilde{\alpha}_{k+1}$ (as a random variable). $p(x)$ is the average bitwise error probability of $\widetilde{\mathbf{y}}_{\mathrm{B}}$ conditioning on $\{\widetilde{\alpha}_{k+1} = x\}$, which is given by \cite[Eq. (13)]{yue2021revisit}
    \begin{equation}   \label{equ::Ana::Pe::px}
         p(x) = \frac{Q(\frac{2x+2}{\sqrt{2N_0}}) }{ Q(\frac{2x+2}{\sqrt{2N_0}}) +  Q(\frac{2x-2}{\sqrt{2N_0}}) } .
     \end{equation}
Then, when $N_0\to 0$, we have \cite{yue2021linear}
     \begin{equation}   
         \lim_{N_0 \to 0}p(x) = \left(1 + \lim\limits_{N_0 \to 0}\exp\left(4x/N_0\right)\right)^{-1}.
     \end{equation}
when $N_0 \to 0$, there are $\widetilde{\alpha}_{k+1} \to 1$ and $f_{\widetilde{\alpha}_{k+1}}(x) \to \delta(x-1)$, where $\delta(x)$ is the Dirac delta function. We then obtain 
\begin{equation} \label{equ::Plist::N0to0}
     \lim_{N_0 \to 0}\mathrm{P_{list}} =  \lim_{N_0 \to 0}\sum_{i=0}^{m} \binom{k}{i} \left(\frac{1}{1 + e^{4/N_0}}\right)^{\!i\!}\left(\frac{e^{4/N_0}}{1 + e^{4/N_0}}\right)^{\!\!k-i}\!\!\!.
\end{equation}

On the other hand, we can derive $\mathrm{P'_{list}}$ as
 \begin{equation}  \label{equ::Ana::Plist'}
     \mathrm{P'_{list}} = \sum_{i=0}^{m'} \binom{k}{i} (p')^{i}(1-p')^{k-i},
 \end{equation}
where $p'$ is the bitwise error probability of $\mathbf{y}_{\mathrm{B}}$. Under AWGN and BPSK, $p'$ is readily given by $p' = Q(\sqrt{2/N_0})$, which also has the following asymptotic property,
\begin{equation}   \label{equ::Ana::P'::app}
    \lim_{N_0 \to 0}p' = \frac{1}{1 + \lim\limits_{N_0 \to 0}\exp\left(\frac{4}{N_0}\right)}.
\end{equation}
Therefore, substituting (\ref{equ::Ana::P'::app}) into (\ref{equ::Ana::Plist'}) and taking $m'=m-1$, we observe that 
\begin{equation} \label{equ::ana::same}
\begin{split}
   \lim_{N_0 \to 0}&\mathrm{P_{list}}- \mathrm{P'_{list}} \\
   &= \lim_{N_0 \to 0}\binom{k}{m} \left(\frac{1}{1 + e^{4/N_0}}\right)^{\!m\!}\left(\frac{e^{4/N_0}}{1 + e^{4/N_0}}\right)^{\!\!k-m} \\
   &\to 0,   
\end{split}
\end{equation}
which indicates that the error performance of Non-GE OSD and standard OSD converges when $N_0$ is small \footnote{We omit the convergence behavior analysis due to the space limit.}. However, when $N_0$ is not negligible, there is $\mathrm{P'_{list}} < \mathrm{P_{list}}$ according to (\ref{equ::ana::same}), since the Non-GE OSD applies TEPs over $\mathbf{y}_{\mathrm{B}}$ rather than MRB $\widetilde{\mathbf{y}}_{\mathrm{B}}$.

In this regard, we define a parameter $\lambda \in (0,1]$ for relaxation and introduce the ``Condition 1'' as follows. If 
 $$\mathrm{P'_{list}} \geq (1-\lambda)\mathrm{P_{list}} $$
is satisfied, we regard that the Non-GE OSD is highly possible to find the correct decoding result, and it should be performed before the standard OSD. According to (\ref{equ::ana::same}), For an arbitrarily small $\lambda$, $\mathrm{P'_{list}} \geq (1-\lambda) \mathrm{P_{list}}$ can always be satisfied at a sufficient large SNR.

In terms of the implementation, $\mathrm{P'_{list}}$ and $\mathrm{P_{list}}$ could be computed prior according to (\ref{equ::Ana::Plist}) and (\ref{equ::Ana::Plist'}) for a given $N_0$. Alternatively, the decoder can perform an adaptive GE reduction based on the quality of instantaneous received signals, i.e., by evaluating $\mathrm{P'_{list}}$ and $\mathrm{P_{list}}$ on the fly. Given the value of $\widetilde{\alpha}_i$ (the random variable of the $i$-th ordered reliability is realized to $\widetilde{\alpha}_i$), the bit-wise error probability of $\widetilde{y}_i$ is derived as
\begin{equation}   \label{equ::Ana::Pi}
    \mathrm{P}(\widetilde{i}) =\frac{1}{1+\exp\left(4\widetilde{\alpha}_i/N_0\right)}.
\end{equation}
Then, the average error probability of MRB, $p$, is evaluated as $p = \frac{1}{k}\sum_{i=1}^{k} \mathrm{P}(\widetilde{i})$,
and $\mathrm{P_{list}}$ is approximately estimated as
\begin{equation}   \label{equ::Ana::Plist::eva}
    \mathrm{P_{list}} \approx \sum_{i=0}^{m} \binom{k}{i} (p)^{i}(1-p)^{k-i}
\end{equation}
It is important to distinguish between (\ref{equ::Ana::Plist::eva}) and (\ref{equ::Ana::Plist}); (\ref{equ::Ana::Plist::eva}) is instantaneous and can vary from block to block in receiving, while (\ref{equ::Ana::Plist}) is theoretically derived based on $N_0$. Similarly, given the value of $\alpha_i$, the bit-wise error probability of $y_i$ is given by $\mathrm{P}(i) =\frac{1}{1+\exp\left(\alpha_i/N_0\right)}$. Then, $\mathrm{P'_{list}}$ can be estimated by substituting $p' = \frac{1}{k}\sum_{i=1}^{k} \mathrm{P}(i)$ into (\ref{equ::Ana::Plist'}).

We note that $\mathrm{P'_{list}}$ and $\mathrm{P_{list}}$ can be computed on the fly very efficiently. The exponential in (\ref{equ::Ana::Pi}) can be implemented by its polynomial approximation or by table lookup. Moreover, $p$ is tiny at high SNRs and (\ref{equ::Ana::Plist::eva}) can be further approximated to $\mathrm{P_{list}}\approx \sum_{i=0}^{m} \binom{k}{i} p^{i}$.

\vspace{-0.25em} 
\subsection{The Second Condition}
\vspace{-0.25em} 
As depicted in Fig. \ref{Fig::structure}, ``Condition 2'' identifies if the Non-GE OSD has found the correct decoding result. ``Condition 2'' should be checked once a codeword is generated during the process of Non-GE \footnote{Nevertheless, it is sufficient to do it only for codewords that can result in a lower WHD than the so-far recorded minimum. See Algorithm \ref{ago::OSDGEreduction}.}, so that the decoding can be terminated early to reduce the complexity.

Leveraging the approach in PB-OSD \cite{yue2021probability}, the decoder can compute the probability that a generated codeword estimate $\mathbf{c}_{\mathbf{e}}$ is the correct decoding result, conditioning on the difference pattern $\mathbf{d}_{\mathbf{e}} = \mathbf{c}_{\mathbf{e}} \oplus \mathbf{y}$. We denote this probability as $\mathrm{Pr}\left(\mathbf{c}_{\mathbf{e}} | \mathbf{d}_{\mathbf{e}} \right)$. Thus, if $\mathrm{Pr}\left(\mathbf{c}_{\mathbf{e}} | \mathbf{d}_{\mathbf{e}} \right)$ is close enough to 1, $\mathbf{c}_{\mathbf{e}}$ is regarded as the correct decoding result. This leads to the ``Condition 2'' as follows. Given a parameter $\tau \in [0,1]$, if 
$$\mathrm{Pr}\left(\mathbf{c}_{\mathbf{e}} | \mathbf{d}_{\mathbf{e}} \right) \geq \tau,$$
$\mathbf{c}_{\mathbf{e}}$ will be claimed as the decoding result and the decoding is terminated. Probability $\mathrm{Pr}\left(\mathbf{c}_{\mathbf{e}} | \mathbf{d}_{\mathbf{e}} \right)$ can be approximated using the approach of \cite[Eq. (3)]{yue2021probability}. Specifically, given the values of ordered reliabilities $\widetilde{\alpha}_i$, we have
    \begin{equation} \label{equ::Psuc::itis}
        \mathrm{Pr}\left(\mathbf{c}_{\mathbf{e}} | \mathbf{d}_{\mathbf{e}} \right) \approx
          			\Bigg(1  +  \frac{(1-\mathrm{P}(\mathbf{e}))2^{k-n}}{\exp\Big(-\frac{4}{N_0}\sum\limits_{d_{\mathbf{e},i} \neq 0}\alpha_i\Big) \prod\limits_{i=1}^{n} (1-\mathrm{P}(i))} \Bigg)^{-1}  ,
    \end{equation}
where $\mathrm{P}(\mathbf{e})$ is the probability that TEP $\mathbf{e}$ can eliminate the errors over $\mathbf{y}_{\mathrm{B}}$, which is derived as
\begin{equation} \label{Equ::Psuc1}
\begin{split}
  \mathrm{P}(\mathbf{e}) & = \prod_{e_i\neq 0} \mathrm{P}(i) \prod_{e_i = 0} (1-\mathrm{P}(i))  \\
  & \overset{(a)}{=} \exp\Big(-\frac{4}{N_0}\sum_{\substack{1 < i \leq k\\d_{\mathbf{e},i} \neq 0}}\alpha_i\Big)\prod_{i=1}^{k} (1-\mathrm{P}(i)) ,
\end{split}
\end{equation}
where step (a) applies the fact $[d_{\mathbf{e}}]_1^k = \mathbf{e}$. We omit the detailed derivation of (\ref{equ::Psuc::itis}), as it can be obtained by applying unordered reliabilities to  \cite[Eq. (2)]{yue2021probability}.

The computation of (\ref{equ::Psuc::itis}) and (\ref{Equ::Psuc1}) is fast. First, the terms $\prod_{i=1}^{n} (1-\mathrm{P}(i))$ and $\prod_{i=1}^{k} (1-\mathrm{P}(i))$ can be reused for different codeword estimates. Then, $\sum_{d_{\mathbf{e},i} \neq 0}\alpha_i$ is exactly the WHD $\mathcal{D}(\mathbf{c}_{\mathbf{e}},\mathbf{y})$, while $\sum_{\substack{1 < i \leq k\\d_{\mathbf{e},i} \neq 0}}\alpha_i$ is a partial WHD $\mathcal{D}([c_{\mathbf{e}}]_1^k,[y]_1^k)$. Considering that $\mathcal{D}(\mathbf{c}_{\mathbf{e}},\mathbf{y})$ is readily computed when generating each codeword estimate (in order to find the best estimate) during the decoding, (\ref{equ::Psuc::itis}) and (\ref{Equ::Psuc1}) are efficiently implemented.
\vspace{-0.3em} 
\subsection{Algorithm}
\vspace{-0.3em} 
We summarize the proposed decoding scheme in Algorithm \ref{ago::OSDGEreduction}. We employ PB-OSD \cite{yue2021probability} to place the standard OSD component (also see Fig. \ref{Fig::structure}). As the latest OSD improvement, PB-OSD shows a significantly reduced complexity compared to the original OSD \cite{Fossorier1995OSD}. To further reduce the possible overhead of the Non-GE OSD, we also integrate the TEPs discarding rule of \cite[Eq. (9)]{yue2021probability} into the Non-GE OSD. Specifically, a probability $\mathrm{P}_p(\mathbf{e})$ \cite[Eq. (8)]{yue2021probability} is computed for TEP $\mathbf{e}$. Then, if $\mathrm{P}_p(\mathbf{e})$ is less than a threshold $\tau_{p}$, the TEP $\mathbf{e}$ is skipped without re-encoding. This discarding rule can be efficiently implemented\footnote{By using a 
monotonicity trick $\mathrm{P}_p(\mathbf{e})$ only needs to be computed for a small amount of TEPs, resulting in a small self-overhead. We refer interested readers to \cite{yue2021probability}.}. Following the approach in \cite{yue2021probability}, $\tau_p$ is set to $\tau_p = 0.002\sqrt{p'/N_{m'}}$ in Algorithm \ref{ago::OSDGEreduction}, where $p'=Q(\sqrt{2/N_0})$ and $N_{m'} = \sum_{i=1}^{m'}\binom{k}{i}$.

If ``Condition 2'' is not satisfied when finishing the Non-GE OSD, the so-far found best estimate $\mathbf{c}_{\mathrm{best}}$ and the recorded minimum WHD will be passed into PB-OSD, which will be executed to find a better codeword estimate than $\mathbf{c}_{\mathrm{best}}$.

\begin{spacing}{1.25}
    \begin{algorithm}
    \small
	\caption{OSD with Adaptive GE Reduction}
	\label{ago::OSDGEreduction}
	\begin{algorithmic} [1]
		\REQUIRE Received signal $\mathbf{r}$, Parameters $\lambda$, $\tau$, and $\tau_p$
		\ENSURE ~Optimal codeword estimate $\hat{\mathbf{c}}_{\mathrm{best}}$

		~\\
		\STATE Initialize $\mathcal{D}_{\min} = +\infty$.
		\STATE Compute $\mathrm{P_{list}}$ and $\mathrm{P'_{list}}$ according to (\ref{equ::Ana::Plist::eva}) and (\ref{equ::Ana::Plist'}), respectively.
		$//$ \textbf{Perform the Non-GE OSD if Condition 1 is true}
		
		\IF{$\mathrm{P'_{list}} \geq (1-\lambda)\mathrm{P_{list}}$} 
		
                \FOR{$i=1:\sum\limits_{j = 0}^{m}\binom{k}{j}$}
        		\STATE Select an unprocessed TEP $\mathbf{e}_i$ with the lowest Hamming weight
        		\STATE Compute $\mathrm{P}_{p}(\mathbf{e})$ according to \cite[Eq. (8)]{yue2021probability}
        		\STATE \textbf{if} $\mathrm{P}_{p}(\mathbf{e})\leq \tau_p$ \textbf{then} \textbf{Continue}
        		\STATE Generate codeword estimate $\mathbf{c}_{\mathbf{e}_i} = \left(\mathbf{y}_{\mathrm{B}}\oplus \mathbf{e}\right)\mathbf{\widetilde G}$
        		\STATE Compute the WHD $\mathcal{D}(\mathbf{c}_{\mathbf{e}_i} , \mathbf{y})$

    		     \IF{$\mathcal{D}(\mathbf{c}_{\mathbf{e}_i} , \mathbf{y}) \leq \mathcal{D}_{\min}$}
        		\STATE Claim $\mathbf{c}_{\mathbf{e}_i}$ as $\mathbf{c}_{\mathrm{best}}$, and  $\mathcal{D}_{\min} \leftarrow \mathcal{D}(\mathbf{c}_{\mathbf{e}_i} , \mathbf{y})$
        		\STATE Calculate $\mathrm{Pr}\left(\mathbf{c}_{\mathbf{e}} | \mathbf{d}_{\mathbf{e}} \right)$ according to  (\ref{equ::Psuc::itis})
        		\STATE \textbf{if} {$\mathrm{Pr}\left(\mathbf{c}_{\mathbf{e}} | \mathbf{d}_{\mathbf{e}} \right) \geq \tau$}
        		    \textbf{then} \textbf{return} $\mathbf{c}_{\mathrm{best}}$ \hspace*{\fill}  $//$\textbf{Condition 2}
        		\ENDIF
        		
    		    \ENDFOR
		\ENDIF
		\STATE Perform PB-OSD \cite{yue2021probability}, with $\mathcal{D}_{\min}$ and $\mathbf{c}_{\mathrm{best}}$ passed in.
	\end{algorithmic}
\end{algorithm} 
\end{spacing}
\vspace{-0.25em} 
\section{Performance and Complexity Consideration} \label{sec::Performance and Complexity}
\vspace{-0.25em} 
\subsubsection{Performance} The proposed decoder will have the same BLER performance as the standard OSD if ``Condition 2'' is effective. Precisely, there will be a decoding error that does not occur in the standard OSD but does occur in the proposed decoder, if and only if "Condition 2" identifies an incorrect estimate as the decoding result. To avoid such decoding errors, the parameter $\tau$ should be chosen with care. Recalling $\mathrm{Pr}\left(\mathbf{c}_{\mathbf{e}} | \mathbf{d}_{\mathbf{e}} \right)$, a large $\tau$ (close to 1) can ensure that $\mathbf{c}_{\mathbf{e}}$ is the correct decoding result with high confidence. In contrast, a small $\tau$ (far from 1) can identify the decoding output earlier while compromising the accuracy. According to our verification, $\tau = 0.95$ can provide a good performance of ``Condition 2'', i.e., identifying correct decoding result accurately while keeping the complexity at a low degree.

\subsubsection{Average Complexity} The average decoding complexity of the proposed decoder can be roughly represented as
\begin{equation} \label{equ::Complexity1}
\begin{split}
    C_{\mathrm{Prop.}} &=  p_1 (C_{\mathrm{NonGE}} + (1-p_2)C_{\mathrm{OSD}}) + (1-p_1)C_{\mathrm{OSD}}\\
    &=  p_1 C_{\mathrm{NonGE}} + (1- p_1p_2)C_{\mathrm{OSD}},
\end{split}
\end{equation}
where $p_1$ and $p_2$ are the probability that ``Condition 1'' and ``Condition 2'' are satisfied, respectively. The complexity $C_{\mathrm{NonGE}}$ is given by $C_{\mathrm{NonGE}} = N_{\mathrm{NonGE}} C_{\mathrm{Re-encoding}}$, where $N_{\mathrm{NonGE}}$ is the average number of TEPs re-encoded in the Non-GE OSD. Therefore, (\ref{equ::Complexity1}) is re-written as
\begin{equation} \label{equ::Complexity2}
\begin{split}
     C_{\mathrm{Prop.}} = & (1- p_1p_2)C_{\mathrm{Preprocessing}} \\
     & + (p_1N_{\mathrm{NonGE}}+(1- p_1p_2)N_a)C_{\mathrm{Re-encoding}}\\
     =& (1- p_1p_2)C_{\mathrm{Preprocessing}} + N_a'C_{\mathrm{Re-encoding}},
\end{split}
\end{equation}
where $N_a' = p_1N_{\mathrm{NonGE}}+(1- p_1p_2)N_a$ is the average number of TEPs in the entire procedure of the proposed decoder. Comparing (\ref{equ::Complexity2}) and (\ref{equ::CmpOSD}), the proposed decoder can have a lower complexity than the standard OSD if $ (1- p_1p_2)<1$ and $N_a' \approx N_a$, i.e., reducing the overhead of preprocessing while maintaining the overhead of reprocessing. 

Rigorously determining $p_1$, $p_2$, and $C_{\mathrm{Prop. }}$ is out of the scope of this paper \footnote{A rigorous $C_{\mathrm{Prop.}}$ shall be determined by considering the correlations 1) between $p_1$ and $p_2$ and 2) between $N_{\mathrm{NonGE}}$ and $p_2$. For example, if $p_1$ is small (a small $\lambda$), $\mathrm{P'_{list}}$ will be large, implying that the Non-GE OSD is likely to find the correct decoding results. As a result, $p_2$ will also be large.}; nevertheless, (\ref{equ::Complexity2}) still provides insights into the parameter selection. A large $\tau$ will decrease $p_2$, and therefore increases $ C_{\mathrm{Prop.}}$. If $\tau$ is assumed to be fixed, a larger $\lambda$ will raise $p_1$, which lowers the preprocessing overhead. Nonetheless, this may increase $N_a'$, hence increasing the cost of reprocessing. We will further demonstrate the impacts of $\lambda$ over $N_a'$ and $(1- p_1p_2)$ via simulations in Section \ref{Sec::Simulation},

\subsubsection{Worst-Case Complexity} The proposed decoder is probable to have a higher worst-case complexity than the standard OSD. However, the increase in the worst-case complexity could be minor. The worst-case complexity of the order-$m$ OSD is given by $ C_{\mathrm{OSD}}^{\mathrm{(worst)}} = C_{\mathrm{Preprocessing}} + N_mC_{\mathrm{Re-encoding}}$, where $N_m$ is the maximum possible number of TEPs, i.e., $N_m = \sum_{i=0}^{m}\binom{k}{m}$. On the other hand, the worst-case decoding of the proposed decoder is derived as
\begin{equation} \label{equ::Complexityworst}
\begin{split}
     C_{\mathrm{Prop.}}^{\mathrm{(worst)}} = & C_{\mathrm{Preprocessing}} + (N_{m'}+N_m)C_{\mathrm{Re-encoding}},\\
\end{split}
\end{equation}
where $N_m'$ is the maximum possible number of TEPs in Non-GE OSD, i.e.,  $N_m' = \sum_{i=0}^{m'}\binom{k}{m}$. Eq. (\ref{equ::Complexityworst}) is obtained by assuming that the Non-GE OSD is always conducted and ``Condition2'' fails for all estimates. Then, by noticing $m' = m-1$ and $k\gg m$, we have $N_m \gg N_m'$, indicating that 
\begin{equation}
    C_{\mathrm{Prop.}}^{\mathrm{(worst)}}\approx C_{\mathrm{OSD}}^{\mathrm{(worst)}}.
\end{equation}
\vspace{-0.25em} 
\section{Simulation Results} \label{Sec::Simulation}
\vspace{-0.25em} 
In this section, we evaluate the performance and complexity of the proposed decoder via simulations. Simulations in this paper were executed on MATLAB 2021a with an i7-10700 CPU. All results were obtained by simulating decoding until 500 decoding errors were collected.
\vspace{-0.25em} 
\subsection{BLER performance}
\begin{figure}  [t]
     \centering
    \begin{tikzpicture}
    \definecolor{applebrown}{rgb}{0.55, 0.71, 0.0}
    \definecolor{english}{rgb}{0.0, 0.5, 0.0}
    \begin{axis}[%
    width=2.8in,
    height=1.4in,
    at={(0.822in,0.529in)},
    scale only axis,
    xmin=-1,
    xmax=8,
    xlabel style={at={(0.5,1ex)},font=\color{white!15!black},font = \small},
    xlabel={SNR [dB]},
    ymin=1e-6,
    ymax=1,
    ymode=log,
    ylabel style={at={(1.5ex,0.5)},font=\color{white!15!black},font = \small},
    ylabel={BLER},
    ytick = {1,1e-1,1e-2,1e-3,1e-4,1e-5},
    yminorticks=true,
    axis background/.style={fill=white},
    tick label style={font=\footnotesize},
    xmajorgrids,
    ymajorgrids,
    yminorgrids,
    major grid style={dotted,black},
    minor grid style={dotted},
    legend style={at={(0,0)}, anchor=south west, legend cell align=left, align=left, draw=white!15!black,font = \tiny	,row sep=-1pt, legend columns=1}
    ]
    
    \addplot [thick, color=black]
          table[row sep=crcr]{%
        10 10\\
        };
    \addlegendentry{The original OSD \cite{Fossorier1995OSD}}
    
    \addplot [ color=red, dashed]
          table[row sep=crcr]{%
        10 10\\
        };
    \addlegendentry{The proposed decoder}

    \addplot [only marks, color=black, mark=square ,mark size = 1.5pt, mark options={solid, black}]
          table[row sep=crcr]{%
        10 10\\
        };
    \addlegendentry{$(64,24)$ eBCH}
    
    \addplot [only marks, color=black, mark=triangle ,mark size = 2.0pt, mark options={solid, black}]
          table[row sep=crcr]{%
        10 10\\
        };
    \addlegendentry{$(64,36)$ eBCH}
    
    \addplot [only marks, color=black, mark=diamond ,mark size = 2.0pt, mark options={solid, black}]
          table[row sep=crcr]{%
        10 10\\
        };
    \addlegendentry{$(64,45)$ eBCH}
    
    \addplot [only marks, color=black, mark=o ,mark size = 2.0pt, mark options={solid, black}]
          table[row sep=crcr]{%
        10 10\\
        };
    \addlegendentry{$(128,106)$ eBCH}

    \addplot [ color=black, mark=square, mark size = 1.54 pt, mark options={solid, black}, forget plot]
    table[row sep=crcr, x expr={\thisrow{SNR}}]{%
                SNR  Cpx\\
    -1.24938736608300	0.228519195612431\\
    -0.849387366083000	0.170706725844998\\
    -0.449387366083000	0.122010736944851\\
    -0.0493873660829993	0.0656512605042017\\
    0.350612633917001	0.0386966953022212\\
    0.750612633917001	0.0195152414035362\\
    1.15061263391700	0.00750187546886722\\
    1.55061263391700	0.00322514061613086\\
    1.95061263391700	0.000990073522859808\\
    2.35061263391700	0.000297017587005362\\
    2.75061263391700	8.12398368964043e-05\\
    3.15061263391700	1.66992524746622e-05\\
    };
    \addlegendentry{$(64,24)$ eBCH}

    \addplot [dashed,color=red, mark=square, mark size = 1.54 pt, mark options={solid, red}, forget plot]
    table[row sep=crcr, x expr={\thisrow{SNR}}]{%
                SNR  Cpx\\
    -0.249387366083000	0.0963081861958267\\
    0.250612633917001	0.0440076279888514\\
    0.750612633917001	0.0182937984023416\\
    1.25061263391700	0.00640396191777313\\
    1.75061263391700	0.00177455990914253\\
    2.25061263391700	0.000451175311686945\\
    2.75061263391700	9.81526585379287e-05\\
    3.25061263391700	1.4035743395750e-05\\
    };
    \addlegendentry{$(64,24)$ eBCH}

    \addplot [color=black, mark=triangle, mark size = 2 pt, mark options={solid, black}, forget plot]
    table[row sep=crcr, x expr={\thisrow{SNR}}]{%
                SNR  Cpx\\
        1.01	0.299043062200957\\
        1.51	0.175438596491228\\
        2.01	0.0968616815187912\\
        2.51	0.0428595919766844\\
        3.01	0.0134220981423816\\
        3.51	0.00394343536315096\\
        4.01	0.000758223694187154\\
        4.51	0.000127519888639432\\
        5.01	1.47450469323048e-05\\
        5.51    1.68e-06\\
    };
    \addlegendentry{$(64,36)$ eBCH}

    \addplot [dashed,color=red, mark=triangle, mark size = 2 pt, mark options={solid, red}, forget plot]
    table[row sep=crcr, x expr={\thisrow{eBN0}+0.5115}]{%
                eBN0  Cpx\\
    1	0.208333333333333\\
    1.50000000000000	0.106685633001422\\
    2	0.0457526307762696\\
    2.50000000000000	0.0149476831091181\\
    3	0.00406157345355591\\
    3.50000000000000	0.000825509339262325\\
    4	0.000131736702427636\\
    4.50000000000000	1.66550191360077e-05\\
    5	1.78000000000000e-06\\
    };
    \addlegendentry{$(64,36)$ eBCH}

    \addplot [color=black, mark=diamond, mark size = 2 , mark options={solid, black}, forget plot]
    table[row sep=crcr, x expr={\thisrow{eBN0}+ 1.4806}]{%
                eBN0  Cpx\\
    0.5	0.549450549450549\\
    1	0.373134328358209\\
    1.5	0.259336099585062\\
    2	0.126198889449773\\
    2.5	0.0598229241445322\\
    3	0.0229568411386593\\
    3.5	0.00712271004871934\\
    4	0.00139593849599449\\
    4.5	0.000290677394600377\\
    5	3.259780940113e-05\\
    5.5	3.89843567403314e-06\\
    };
    \addlegendentry{$(64,45)$ eBCH}

    \addplot [dashed,color=red, mark=diamond, mark size = 2 , mark options={solid, red}, forget plot]
    table[row sep=crcr, x expr={\thisrow{eBN0}+ 1.4806}]{%
                eBN0  Cpx\\
    1	0.390625000000000\\
    1.5	0.257289879931389\\
    2	0.131694468832309\\
    2.5	0.0594059405940594\\
    3	0.0242502627111794\\
    3.5	0.00674930819590992\\
    4	0.00159797164132994\\
    4.5	0.000275109975212591\\
    5	4.41087409850923e-05\\
    5.5	4.69277717458210e-06\\
    };
    \addlegendentry{$(64,45)$ eBCH}

    \addplot [color=black, mark=o ,mark size = 2pt, mark options={solid, black}, forget plot]
      table[row sep=crcr]{%
    2.19125891280883	0.97799511002445\\
    2.69125891280883	0.934579439252336\\
    3.19125891280883	0.849256900212314\\
    3.69125891280883	0.683760683760684\\
    4.19125891280883	0.471142520612485\\
    4.69125891280883	0.236127508854782\\
    5.19125891280883	0.0888296691094826\\
    5.69125891280883	0.0281729821101564\\
    6.19125891280883	0.00505439795802323\\
    6.69125891280883	0.000686511423550088\\
    7.19125891280883	6.36064849577776e-05\\
    7.49125891280883	1.211e-05\\
    };
    \addlegendentry{$(128,106)$ eBCH}
    
    \addplot [dashed,color=red, mark=o ,mark size = 2pt, mark options={solid, red}, forget plot]
      table[row sep=crcr]{%
    3.19125891280883	0.87977 \\
    3.69125891280883	0.66815\\
    4.19125891280883	0.42433\\
    4.69125891280883	0.24773\\
    5.19125891280883	0.09446\\
    5.69125891280883	0.03010\\
    6.19125891280883	0.00530\\
    6.69125891280883    0.00077\\
    7.19125891280883	0.00007\\
    7.69125891280883	4.811e-06\\
    };
    \addlegendentry{$(128,106)$ eBCH}

    \end{axis}
    \end{tikzpicture}%
	\vspace{-0.5em}
    \caption{BLER performance of the proposed decoder and the original OSD \cite{Fossorier1995OSD}. The decoding order is optimal, i.e., $m = \lceil d_{\mathrm{H}}/4-1\rceil$. }
    \vspace{-0.25em}
	\label{Fig::Performance}
        
	\end{figure}
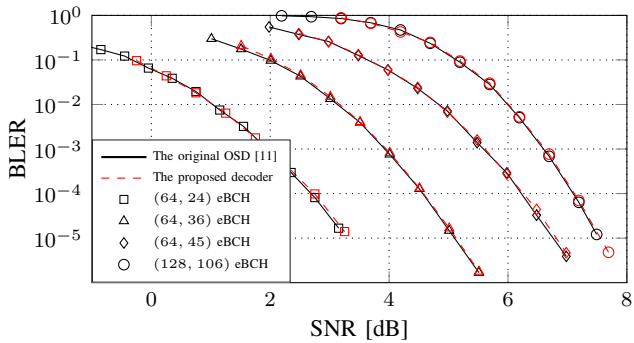

\vspace{-0.25em} 
We demonstrate the BLER performance of the proposed decoder in decoding various length-64 and length-128 eBCH codes in Fig \ref{Fig::Performance}. Parameters are set to $\tau = 0.95$ and $\lambda = 0.05$. The proposed decoder is compared with the original OSD \cite{Fossorier1995OSD}. The decoding order of both decoders are set to $m = \lceil d_{\mathrm{H}}/4-1\rceil$. The Non-GE part in the proposed decoder accordingly has the order of $m-1$. It can be seen that the proposed decoder has nearly the same BLER as the original OSD. In other words, the proposed decoder can approach the near-ML performance (recall that OSD at the order $m = \lceil d_{\mathrm{H}}/4-1\rceil$ is ML \cite{Fossorier1995OSD}).

\vspace{-0.25em} 
\subsection{Computational Complexity}
\vspace{-0.25em} 
We evaluate the complexity from two aspects; 1) the average decoding time, and 2) the number of average TEPs in decoding a single codeword. The average decoding time was measured in the same MATLAB environments for all decoders, reflecting the overall decoding computational complexity. On the other hand, according to (\ref{equ::CmpOSD}), the number of average TEPs can reflect the complexity of OSD-based decoders when TEPs are many, i.e., when reprocessing dominates the decoding complexity rather than preprocessing. 

We compare the proposed decoder with the original OSD \cite{Fossorier1995OSD}, the latest PB-OSD \cite{yue2021probability}, as well as the recent approach proposed by Choi \emph{et. al.} \cite{choi2021fast}. We implemented Choi's approach with three offline permuted generator matrices as per the implementation in \cite{choi2021fast}. Decoding the $(64,36)$ eBCH code, we compare the decoding time and numbers of TEPs in Fig. \ref{Fig::64-36-Time} and Fig. \ref{Fig::64-36-Na}, respectively. Parameters of the proposed decoder are set to $\tau = 0.95$ and $\lambda = 0.05$. As shown, PB-OSD has a significantly reduced complexity compared to the original OSD; however, it shows a complexity floor at high SNRs due to the overhead of GE. We highlight that this complexity floor is broken by the proposed decoder, which significantly reduces the decoding complexity of PB-OSD at high SNRs. We note that the proposed algorithm slightly increases the number of TEPs at high SNRs compared to PB-OSD; nevertheless, this is acceptable as the reduction of GE results in a lower overall decoding time. Choi's approach can also break the complexity floor. However, it may have a significantly increased number of TEPs at low SNRs compared to PB-OSD or even the original OSD. This is because it performs additional reprocessing over multiple generator matrices\footnote{Although the decoding time of Choi's approach is high at low SNRs in Fig. \ref{Fig::64-36-Time}, \cite{choi2021fast} indicated that the decoding efficiency can be improved by employing parallelism; nevertheless, decoders in this paper were not implemented with parallel structures. We refer interested readers to \cite{choi2021fast} for details.}. In contrast, the proposed algorithm introduces virtually no additional TEPs and decoding complexity at low SNRs compared to PB-OSD.

We further depict the average decoding time and number of TEPs in decoding the $(128,106)$ eBCH code in Fig. \ref{Fig::128-106}. Breaking the complexity floor is more critical for such high-rate codes, because 1) OSD usually requires low orders to decode high-rate codes and therefore the number of required TEPs is not large, and 2) the overhead of GE is higher for generator matrices with more rows (a larger $k$). As shown by Fig. \ref{Fig::128-106}, the proposed decoder has a significantly low complexity in decoding $(128,106)$ eBCH code at high SNRs. For example, at the SNR of 9 dB, the proposed decoder requires 0.09 ms to decode a codeword, compared to 1.9 ms of PB-OSD suffering from the complexity floor.

           \begin{figure}
             \centering
             \vspace{-0.25em}
             \hspace{-0.81em}
             \begin{subfigure}[b]{0.49\columnwidth}
                 \centering
                \begin{tikzpicture}
                \begin{axis}[%
                width=1.35in,
                height=1.5in,
                at={(0.785in,0.587in)},
                scale only axis,
                xmin=2.5,
                xmax=8.5,
                xlabel style={at={(0.5,2ex)},font=\color{white!15!black},font=\scriptsize},
                xlabel={SNR (dB)},
                ymode=log,
                ymin=1e-2,
                ymax=100,
                yminorticks=true,
                ylabel style={at={(3ex,0.5)},font=\color{white!15!black},font=\scriptsize},
                ylabel={Average Decoding Time (ms)},
                axis background/.style={fill=white},
                tick label style={font=\tiny},
                xmajorgrids,
                ymajorgrids,
                yminorgrids,
                minor grid style={dotted},
                major grid style={dotted,black},
                legend style={at={(0,0)}, anchor=south west, legend cell align=left, align=left, draw=white!15!black,font = \tiny,row sep=-3pt}
                ]
                
                \addplot [color=black, mark=square ,mark size = 1.5pt, mark options={solid, black}]
                  table[row sep=crcr]{%
                2.5  17.207   \\
                3	17.207  \\
                3.5	17.207  \\
                4	17.207  \\
                4.5	17.207  \\
                5	17.207  \\
                5.5	17.207  \\
                6	17.207  \\
                6.5	17.207  \\
                7	17.207  \\
                7.5	17.207  \\
                8	17.207  \\
                8.5	17.207  \\
                };

                \addplot [color=blue, mark=o ,mark size = 2pt, mark options={solid, blue}]
                  table[row sep=crcr]{%
                2.5000    0.4939\\
                3.0000    0.3067\\
                3.5000    0.1933\\
                4.0000    0.1283\\
                4.5000    0.1030\\
                5.0000    0.0926\\
                5.5000    0.0903\\
                6.0000    0.0895\\
                6.5000    0.0895\\
                7.0000    0.0904\\
                7.5000    0.0912\\
                8.0000    0.0929\\
                8.5000    0.0906\\
                };
                
                \addplot [color=red, mark=x ,mark size = 2pt, mark options={solid, red}]
                  table[row sep=crcr]{%
                2.5  0.529  \\
                3	0.299472683393288\\
                3.5	0.193465876192601\\
                4	0.131855751000000\\
                4.5	0.104433565000000\\
                5	0.0891703520000000\\
                5.5	0.0761599070000000\\
                6	0.0625690520000000\\
                6.5	0.0477919320000000\\
                7	0.0373623350000000\\
                7.5	0.0300554940000000\\
                8	0.0261969560000000\\
                8.5	0.0244859210000000\\
                };
                
                \addplot [color=green, mark=diamond ,mark size = 2pt, mark options={solid, green}]
                  table[row sep=crcr]{%
                2.5000   23.3453\\
                3.0000   19.1801\\
                3.5000   15.8998\\
                4.0000   11.8227\\
                4.5000    7.1446\\
                5.0000    4.4865\\
                5.5000    2.3678\\
                6.0000    1.2179\\
                6.5000    0.3403\\
                7.0000    0.1016\\
                7.5000    0.0540\\
                8.0000    0.0376\\
                8.5000    0.0347\\
                };
                
        
                \end{axis}
                \end{tikzpicture}%
                \vspace{-0.3em}
                \caption{Average Decoding Time}  
                \vspace{-0.3em}
                \label{Fig::64-36-Time}

             \end{subfigure}
             \hspace{-0.4em}
             \begin{subfigure}[b]{0.49\columnwidth}
                \centering
                \begin{tikzpicture}
                \begin{axis}[%
                width=1.35in,
                height=1.5in,
                at={(0.642in,0.505in)},
                scale only axis,
                xmin=2.5,
                xmax=8.5,
                xlabel style={at={(0.5,2ex)},font=\color{white!15!black},font=\scriptsize},
                xlabel={SNR (dB)},
                ymode=log,
                ymin=0,
                ymax=1e5,
                ylabel style={at={(4ex,0.5)},font=\color{white!15!black},font=\scriptsize},
                ylabel={Average number of TEPs},
                axis background/.style={fill=white},
                tick label style={font=\tiny},
                xmajorgrids,
                ymajorgrids,
                yminorgrids,
                minor grid style={dotted},
                major grid style={dotted,black},
                legend style={at={(1,1.03)}, anchor=south east, legend cell align=left, align=left, draw=white!15!black,font = \tiny,row sep=-3pt, legend columns=2}
                ]

                \addplot [color=black, mark=square ,mark size = 1.5pt, mark options={solid, black}]
                      table[row sep=crcr]{%
                    10 10\\
                    };
                \addlegendentry{Original OSD \cite{Fossorier1995OSD}}
                
                \addplot  [color=blue, mark=o ,mark size = 2pt, mark options={solid, blue}]
                      table[row sep=crcr]{%
                    10 10\\
                    };
                \addlegendentry{PB-OSD \cite{yue2021probability}}
                
                \addplot  [color=green, mark=diamond ,mark size = 2pt, mark options={solid, green}]
                      table[row sep=crcr]{%
                    10 10\\
                    };
                \addlegendentry{Choi's \cite{choi2021fast}}
                
                \addplot [color=red, mark=x ,mark size = 2pt, mark options={solid, red}]
                      table[row sep=crcr]{%
                    10 10\\
                    };
                \addlegendentry{Proposed}
                
                \addplot [color=red, mark=x ,mark size = 2pt, mark options={solid, red}, forget plot]
                  table[row sep=crcr]{%
                2.5000  237.7881\\
                3.0000  126.0819\\
                3.5000   60.1544\\
                4.0000   21.7310\\
                4.5000    8.1435\\
                5.0000    5.6081\\
                5.5000    6.1709\\
                6.0000    6.4867\\
                6.5000    6.2388\\
                7.0000    5.1092\\
                7.5000    3.3950\\
                8.0000    2.2321\\
                8.5000    1.5305\\
                };
                \addlegendentry{The proposed}
                
                \addplot [color=blue, mark=o ,mark size = 2pt, mark options={solid, blue}, forget plot]
                  table[row sep=crcr]{%
                    2.5000  227.3613\\
                    3.0000  127.2286\\
                    3.5000   60.0300\\
                    4.0000   21.1256\\
                    4.5000    7.2184\\
                    5.0000    2.4126\\
                    5.5000    1.3807\\
                    6.0000    1.0722\\
                    6.5000    1.0136\\
                    7.0000    1.0094\\
                    7.5000    1.0019\\
                    8.0000    1.0003\\
                    8.5000    1.0000\\
                };
                \addlegendentry{PB-OSD \cite{yue2021probability}}
                
                \addplot [color=green, mark=diamond ,mark size = 2pt, mark options={solid, green}, forget plot]
                  table[row sep=crcr]{%
                2.5	18560.8650000000\\
                3	15360.3655000000\\
                3.5	12126.3660000000\\
                4	9369.74250000000\\
                4.5	5870.60000000000\\
                5	3666.04900000000\\
                5.5	1858.25800000000\\
                6	852.211500000000\\
                6.5	250.959000000000\\
                7	55.7085000000000\\
                7.5	17.3245000000000\\
                8	4.28850000000000\\
                8.5	2.01350000000000\\
                };
                
                \addplot [color=black, mark=square ,mark size = 1.5pt, mark options={solid, black}, forget plot]
                  table[row sep=crcr]{%
                2.5  7807\\
                3	  7807\\
                3.5	  7807\\
                4	  7807\\
                4.5	  7807\\
                5	  7807\\
                5.5	  7807\\
                6	  7807\\
                6.5	  7807\\
                7	  7807\\
                7.5	  7807\\
                8     7807\\
                8.5	  7807\\
                };

                \end{axis}
                \end{tikzpicture}%
                \vspace{-0.3em}
                 \caption{Average Number of TEPs}
                 \vspace{-0.3em}
                 \label{Fig::64-36-Na}
             \end{subfigure}
             
              \vspace{-0.11em}
             \caption{The average decoding time and number of TEPs of various decoders in decoding the$(64,36)$ eBCH code with order 3.}
            \vspace{-0.61em}
             \label{Fig::64-36}
        \end{figure}
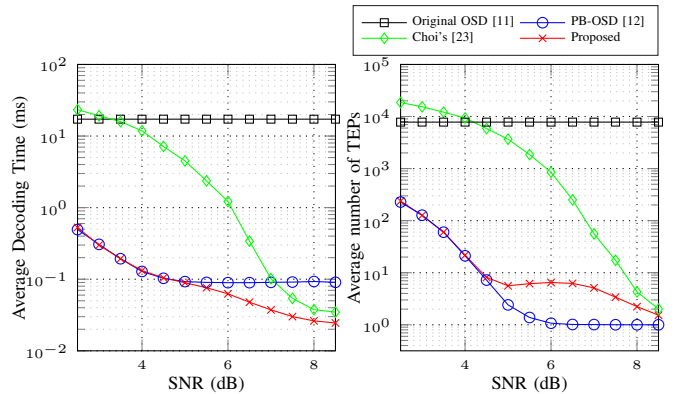

           \begin{figure}
             \centering
             \vspace{-0.25em}
             \hspace{-0.81em}
             \begin{subfigure}[b]{0.49\columnwidth}
                 \centering
                \begin{tikzpicture}
                \begin{axis}[%
                width=1.35in,
                height=1.5in,
                at={(0.785in,0.587in)},
                scale only axis,
                xmin= 4,
                xmax= 10,
                xlabel style={at={(0.5,2ex)},font=\color{white!15!black},font=\scriptsize},
                xlabel={SNR (dB)},
                ymode=log,
                ymin=1e-2,
                ymax=100,
                yminorticks=true,
                ylabel style={at={(3ex,0.5)},font=\color{white!15!black},font=\scriptsize},
                ylabel={Average Decoding Time (ms)},
                axis background/.style={fill=white},
                tick label style={font=\tiny},
                xmajorgrids,
                ymajorgrids,
                yminorgrids,
                minor grid style={dotted},
                major grid style={dotted,black},
                legend style={at={(0,0)}, anchor=south west, legend cell align=left, align=left, draw=white!15!black,font = \tiny,row sep=-3pt}
                ]
                
                \addplot [color=black, mark=square ,mark size = 1.5pt, mark options={solid, black}]
                  table[row sep=crcr]{%
                4	27.268 \\
                4.5	27.268 \\
                5	27.268 \\
                5.5	27.268 \\
                6	27.268 \\
                6.5	27.268 \\
                7	27.268 \\
                7.5	27.268 \\
                8	27.268 \\
                8.5	27.268 \\
                9	27.268 \\
                9.5	27.268 \\
                10	27.268 \\
                };

                \addplot [color=blue, mark=o ,mark size = 2pt, mark options={solid, blue}]
                  table[row sep=crcr]{%
                4.0000    3.7135\\
                4.5000    3.2725\\
                5.0000    2.7513\\
                5.5000    2.1758\\
                6.0000    1.9852\\
                6.5000    1.8191\\
                7.0000    1.8289\\
                7.5000    1.8615\\
                8.0000    1.7834\\
                8.5000    1.8478\\
                9.0000    1.8499\\
                9.5000    1.8798\\
               10.0000    1.9112\\
                };
                
                \addplot [color=red, mark=x ,mark size = 2pt, mark options={solid, red}]
                  table[row sep=crcr]{%
                4.0000    3.5236\\
                4.5000    3.3048\\
                5.0000    2.7370\\
                5.5000    2.3815\\
                6.0000    2.0316\\
                6.5000    1.8902\\
                7.0000    1.7499\\
                7.5000    1.3460\\
                8.0000    0.8306\\
                8.5000    0.4576\\
                9.0000    0.1997\\
                9.5000    0.0908\\
               10.0000    0.0599\\
                };
                
                \addplot [color=green, mark=diamond ,mark size = 2pt, mark options={solid, green}]
                  table[row sep=crcr]{%
                4.0000   75.0353\\
                4.5000   70.8606\\
                5.0000   74.0493\\
                5.5000   61.5666\\
                6.0000   54.5038\\
                6.5000   39.9972\\
                7.0000   25.6771\\
                7.5000   14.4564\\
                8.0000    6.9522\\
                8.5000    3.4117\\
                9.0000    1.3649\\
                9.5000    0.6544\\
               10.0000    0.2916\\
                };
                

                \end{axis}
                \end{tikzpicture}%
                \vspace{-0.3em}
                \caption{Average Decoding Time}  
                \vspace{-0.3em}
                \label{Fig::128-106-Time}

             \end{subfigure}
             \hspace{-0.4em}
             \begin{subfigure}[b]{0.49\columnwidth}
                \centering
                \begin{tikzpicture}
                \begin{axis}[%
                width=1.35in,
                height=1.5in,
                at={(0.642in,0.505in)},
                scale only axis,
                xmin=4,
                xmax=10,
                xlabel style={at={(0.5,2ex)},font=\color{white!15!black},font=\scriptsize},
                xlabel={SNR (dB)},
                ymode=log,
                ymin=0,
                ymax=1e5,
                ylabel style={at={(4ex,0.5)},font=\color{white!15!black},font=\scriptsize},
                ylabel={Average number of TEPs},
                axis background/.style={fill=white},
                tick label style={font=\tiny},
                xmajorgrids,
                ymajorgrids,
                yminorgrids,
                minor grid style={dotted},
                major grid style={dotted,black},
                legend style={at={(1,1.03)}, anchor=south east, legend cell align=left, align=left, draw=white!15!black,font = \tiny,row sep=-3pt, legend columns=2}
                ]

                \addplot [color=black, mark=square ,mark size = 1.5pt, mark options={solid, black}]
                      table[row sep=crcr]{%
                    100 100\\
                    };
                \addlegendentry{Original OSD \cite{Fossorier1995OSD}}
                
                \addplot  [color=blue, mark=o ,mark size = 2pt, mark options={solid, blue}]
                      table[row sep=crcr]{%
                    100 100\\
                    };
                \addlegendentry{PB-OSD \cite{yue2021probability}}
                
                \addplot  [color=green, mark=diamond ,mark size = 2pt, mark options={solid, green}]
                      table[row sep=crcr]{%
                    100 100\\
                    };
                \addlegendentry{Choi's \cite{choi2021fast}}
                
                \addplot [color=red, mark=x ,mark size = 2pt, mark options={solid, red}]
                      table[row sep=crcr]{%
                    100 100\\
                    };
                \addlegendentry{Proposed}
                
                \addplot [color=red, mark=x ,mark size = 2pt, mark options={solid, red}, forget plot]
                  table[row sep=crcr]{%
                     4.0000  506.3795\\
                    4.5000  434.6542\\
                    5.0000  310.6941\\
                    5.5000  148.9069\\
                    6.0000   49.2371\\
                    6.5000   13.7714\\
                    7.0000    5.2818\\
                    7.5000    4.4104\\
                    8.0000    4.5196\\
                    8.5000    4.3223\\
                    9.0000    2.8542\\
                    9.5000    1.8470\\
                   10.0000    1.3805\\
                };
                \addlegendentry{The proposed}
                
                \addplot [color=blue, mark=o ,mark size = 2pt, mark options={solid, blue}, forget plot]
                  table[row sep=crcr]{%
                    4.0000  560.9407\\
                    4.5000  447.0570\\
                    5.0000  307.8416\\
                    5.5000  145.6763\\
                    6.0000   50.1076\\
                    6.5000   14.0725\\
                    7.0000    3.7646\\
                    7.5000    1.3143\\
                    8.0000    1.0450\\
                    8.5000    1.0067\\
                    9.0000    1.0049\\
                    9.5000    1.0000\\
                   10.0000    1.0000\\
                };
                \addlegendentry{PB-OSD \cite{yue2021probability}}
                
                \addplot [color=green, mark=diamond ,mark size = 2pt, mark options={solid, green}, forget plot]
                  table[row sep=crcr]{%
                    4	20683.2699228792\\
                    4.5	19378.4144144144\\
                    5	16969.3082489146\\
                    5.5	14220.3410000000\\
                    6	11730.3995000000\\
                    6.5	8581.93050000000\\
                    7	5441.44050000000\\
                    7.5	3374.11000000000\\
                    8	1697.45850000000\\
                    8.5	819.381000000000\\
                    9	298.622000000000\\
                    9.5	145.275500000000\\
                    10	59.0740000000000\\
                };
                
                \addplot [color=black, mark=square ,mark size = 1.5pt, mark options={solid, black}, forget plot]
                  table[row sep=crcr]{%
                4	  5672\\
                4.5	  5672\\
                5	  5672\\
                5.5	  5672\\
                6	  5672\\
                6.5	  5672\\
                7	  5672\\
                7.5	  5672\\
                8     5672\\
                8.5	  5672\\
                9	  5672\\
                9.5	  5672\\
                10	  5672\\
                };

                \end{axis}
                \end{tikzpicture}%
                \vspace{-0.3em}
                 \caption{Number of decoding iterations}
                 \vspace{-0.3em}
                 \label{Fig::128-106-Na}
             \end{subfigure}
             
              \vspace{-0.11em}
             \caption{The average decoding time and number of TEPs of various decoders in decoding the $(128,106)$ eBCH code with order 2.}
            \vspace{-0.61em}
             \label{Fig::128-106}
        \end{figure}

\vspace{-0.25em} 
\subsection{Impacts of the Parameter $\lambda$}
\vspace{-0.25em} 
In this section, we investigate the impact of $\lambda$ on the number of TEPs, i.e., $N_a'$, and the probability of GE reduction, i.e., $p_1p_2$ in (\ref{equ::Complexity2}). In Fig \ref{Fig::64-36-PARA}. we illustrate $N_a'$ and $p_1p_2$ at different selections of $\lambda$ in decoding the $(64,36)$ eBCH code. As shown, the proposed approach can almost reduce all GEs from OSD at high SNRs. Furthermore, it can be seen that there is a trade-off in selecting $\lambda$. Specifically, if $\lambda$ is large, the decoder can reduce more GEs at low SNRs at the expense of increasing the number of TEPs. On the other hand, a smaller $\lambda$ results in a smaller number of TEPs, but also comprises the performance of GE reduction. We identify that the optimal selection of $\lambda$ should be found by further research. Nevertheless, from our simulations, $\lambda = 0.05$ could provide a good trade-off between $N_a'$ and $p_1p_2$, and result in a low decoding complexity.

    \begin{figure}
             \centering
             \vspace{-0.25em}
             \hspace{-0.81em}
             \begin{subfigure}[b]{0.49\columnwidth}
                 \centering
                \begin{tikzpicture}
                \begin{axis}[%
                width=1.35in,
                height=1.5in,
                at={(0.785in,0.587in)},
                scale only axis,
                xmin=2.5,
                xmax=8.5,
                xlabel style={at={(0.5,2ex)},font=\color{white!15!black},font=\scriptsize},
                xlabel={SNR (dB)},
                ymin=0,
                ymax=1,
                yminorticks=true,
                ylabel style={at={(3ex,0.5)},font=\color{white!15!black},font=\scriptsize},
                ylabel={Probability of GE Reduction, $p_1p_2$},
                axis background/.style={fill=white},
                tick label style={font=\tiny},
                xmajorgrids,
                ymajorgrids,
                yminorgrids,
                minor grid style={dotted},
                major grid style={dotted,black},
                legend style={at={(0,0)}, anchor=south west, legend cell align=left, align=left, draw=white!15!black,font = \tiny,row sep=-3pt}
                ]
                
                \addplot [color=black, mark=square ,mark size = 1.5pt, mark options={solid, black}]
                  table[row sep=crcr]{%
                    2.5000    0.1242\\
                    3.0000    0.2599\\
                    3.5000    0.4318\\
                    4.0000    0.5938\\
                    4.5000    0.7425\\
                    5.0000    0.8433\\
                    5.5000    0.9101\\
                    6.0000    0.9496\\
                    6.5000    0.9767\\
                    7.0000    0.9899\\
                    7.5000    0.9960\\
                    8.0000    0.9986\\
                    8.5000    0.9996\\
                };

                \addplot [color=black, mark=o ,mark size = 2pt, mark options={solid, black}]
                  table[row sep=crcr]{%
                2.5000    0.0055\\
                3.0000    0.0225\\
                3.5000    0.0814\\
                4.0000    0.2026\\
                4.5000    0.3931\\
                5.0000    0.6114\\
                5.5000    0.7881\\
                6.0000    0.9033\\
                6.5000    0.9618\\
                7.0000    0.9860\\
                7.5000    0.9956\\
                8.0000    0.9983\\
                8.5000    0.9993\\
                };
                
                \addplot [color=black, mark=x ,mark size = 2pt, mark options={solid, black}]
                  table[row sep=crcr]{%
                2.5000         0\\
                3.0000    0.0028\\
                3.5000    0.0155\\
                4.0000    0.0588\\
                4.5000    0.1636\\
                5.0000    0.3452\\
                5.5000    0.5561\\
                6.0000    0.7493\\
                6.5000    0.8853\\
                7.0000    0.9572\\
                7.5000    0.9853\\
                8.0000    0.9947\\
                8.5000    0.9991\\
                };
                
                \addplot [color=black, mark=diamond ,mark size = 2pt, mark options={solid, black}]
                  table[row sep=crcr]{%
                2.5000         0\\
                3.0000    0.0004\\
                3.5000    0.0032\\
                4.0000    0.0154\\
                4.5000    0.0605\\
                5.0000    0.1613\\
                5.5000    0.3378\\
                6.0000    0.5538\\
                6.5000    0.7407\\
                7.0000    0.8719\\
                7.5000    0.9430\\
                8.0000    0.9792\\
                8.5000    0.9933\\
                };

                \end{axis}
                \end{tikzpicture}%
                \vspace{-0.3em}
                \caption{Probability of GE reduction, $p_1p_2$}  
                \vspace{-0.3em}
                \label{Fig::64-36-p1p2}

             \end{subfigure}
             \hspace{-0.4em}
             \begin{subfigure}[b]{0.49\columnwidth}
                \centering
                \begin{tikzpicture}
                \begin{axis}[%
                width=1.35in,
                height=1.5in,
                at={(0.642in,0.505in)},
                scale only axis,
                xmin=2.5,
                xmax=8.5,
                xlabel style={at={(0.5,2ex)},font=\color{white!15!black},font=\scriptsize},
                xlabel={SNR (dB)},
                ymode=log,
                ymin=0,
                ymax=1e3,
                ylabel style={at={(4ex,0.5)},font=\color{white!15!black},font=\scriptsize},
                ylabel={Average number of TEPs, $N_a'$},
                axis background/.style={fill=white},
                tick label style={font=\tiny},
                xmajorgrids,
                ymajorgrids,
                yminorgrids,
                minor grid style={dotted},
                major grid style={dotted,black},
                legend style={at={(1,1.03)}, anchor=south east, legend cell align=left, align=left, draw=white!15!black,font = \tiny,row sep=-3pt, legend columns=2}
                ]

                \addplot [only marks , color=black, mark=square ,mark size = 1.5pt, mark options={solid, black}]
                      table[row sep=crcr]{%
                    10 10\\
                    };
                \addlegendentry{$\lambda = 0.5$}
                
                \addplot  [only marks , color=black, mark=o ,mark size = 2pt, mark options={solid, black}]
                      table[row sep=crcr]{%
                    10 10\\
                    };
                \addlegendentry{$\lambda = 0.2$}
                
                \addplot  [only marks ,  color=black, mark=x ,mark size = 2pt, mark options={solid, black}]
                      table[row sep=crcr]{%
                    10 10\\
                    };
                \addlegendentry{$\lambda = 0.1$}
                
                \addplot [only marks , color=black, mark=diamond ,mark size = 2pt, mark options={solid, black}]
                      table[row sep=crcr]{%
                    10 10\\
                    };
                \addlegendentry{$\lambda = 0.05$}
                
                \addplot [color=black, mark=square ,mark size = 1.5pt, mark options={solid, black}, forget plot]
                  table[row sep=crcr]{%
                2.5000  286.2038\\
                3.0000  234.2912\\
                3.5000  210.4110\\
                4.0000  194.6784\\
                4.5000  153.7456\\
                5.0000  110.8260\\
                5.5000   71.8238\\
                6.0000   43.2638\\
                6.5000   22.5609\\
                7.0000   11.2657\\
                7.5000    5.6918\\
                8.0000    2.9838\\
                8.5000    1.7554\\
                };
                \addlegendentry{$\lambda = 0.5$}
                
                \addplot [dashed, color=black, forget plot]
                  table[row sep=crcr]{%
                    2.5000  227.3613\\
                    3.0000  127.2286\\
                    3.5000   60.0300\\
                    4.0000   21.1256\\
                    4.5000    7.2184\\
                    5.0000    2.4126\\
                    5.5000    1.3807\\
                    6.0000    1.0722\\
                    6.5000    1.0136\\
                    7.0000    1.0094\\
                    7.5000    1.0019\\
                    8.0000    1.0003\\
                    8.5000    1.0000\\
                };
                \addlegendentry{PB-OSD \cite{yue2021probability}}
                
                \addplot [color=black, mark=o ,mark size = 2pt, mark options={solid, black}, forget plot]
                  table[row sep=crcr]{%
                    2.5000  231.0426\\
                    3.0000  133.1721\\
                    3.5000   67.0858\\
                    4.0000   45.4232\\
                    4.5000   44.7203\\
                    5.0000   46.5570\\
                    5.5000   43.1711\\
                    6.0000   31.1091\\
                    6.5000   20.3303\\
                    7.0000   10.4390\\
                    7.5000    5.3479\\
                    8.0000    3.1065\\
                    8.5000    1.9397\\
                };
                
                \addplot [color=black, mark=x ,mark size = 2pt, mark options={solid, black}, forget plot]
                  table[row sep=crcr]{%
                    2.5000  236.2947\\
                    3.0000  126.9369\\
                    3.5000   64.0263\\
                    4.0000   25.0845\\
                    4.5000   16.4156\\
                    5.0000   15.9237\\
                    5.5000   16.9914\\
                    6.0000   16.6750\\
                    6.5000   12.7273\\
                    7.0000    7.9841\\
                    7.5000    4.5692\\
                    8.0000    2.8716\\
                    8.5000    1.6094\\
                };

                  \addplot [color=black, mark=diamond ,mark size = 2pt, mark options={solid, black}, forget plot]
                  table[row sep=crcr]{%
                    2.5000  228.1837\\
                    3.0000  127.0546\\
                    3.5000   59.8820\\
                    4.0000   22.7882\\
                    4.5000    8.0843\\
                    5.0000    5.7473\\
                    5.5000    6.4787\\
                    6.0000    6.9854\\
                    6.5000    6.2725\\
                    7.0000    5.0019\\
                    7.5000    3.5111\\
                    8.0000    2.2742\\
                    8.5000    1.6317\\
                };

                \draw[->,black] (axis cs: 4.8,1.5) -- (axis cs: 5.25,1.5);
                \node[anchor=south] at (axis cs: 4,1){\scriptsize PB-OSD};

                \end{axis}
                \end{tikzpicture}%
                \vspace{-0.3em}
                 \caption{Average number of TEPs, $N_a'$}
                 \vspace{-0.3em}
                 \label{Fig::64-36-N'}
             \end{subfigure}
             
              \vspace{-0.11em}
             \caption{The impacts of values of $\lambda$ in decoding the $(64,36)$ eBCH code with order 3.}
            \vspace{-0.61em}
             \label{Fig::64-36-PARA}
        \end{figure}

\vspace{-0.3em}
\section{Conclusion} \label{sec::Conclusion}
\vspace{-0.3em}

In this paper, we designed an efficient ordered-statistics decoding (OSD) algorithm with adaptive GE reduction. The proposed decoder employs two conditions. The first condition decides whether to conduct the OSD decoding without Gaussian elimination (GE). If the OSD without GE is performed, the second condition identifies if the correct decoding result has been found in the Non-GE decoding process. If so, the standard OSD with GE can be avoided. The proposed decoding algorithm is an effective solution to the ``complexity floor'' owning to the overhead of GE in OSD decoders. Simulation results indicated that compared to the approaches from the literature, the proposed decoder can significantly reduce the decoding complexity at high SNRs while maintaining the error performance of the original OSD.






%
	\vspace{-0.3em}

\bibliography{BibAbrv/IEEEabrv, BibAbrv/OSDAbrv, BibAbrv/SurveyAbrv, BibAbrv/ClassicAbrv, BibAbrv/MathAbrv,BibAbrv/NOMAAbrv, BibAbrv/GrandAbrv}
\bibliographystyle{IEEEtran}

	\vspace{-0.3em}

\end{document}